\documentclass[12pt]{article}
\usepackage{amsfonts,amssymb,amsthm}

\def\be{\begin{equation}}
\def\ee{\end{equation}}
\def\bea{\begin{eqnarray}}
\def\eea{\end{eqnarray}}

\def\R{{\mathbb R}}

\begin{document}
\thispagestyle{empty}
\vspace{2.5cm}

\begin{center}
\bf{\Large The  Drinfeld double $gl(n)\oplus t_n$}
\end{center}

\bigskip\bigskip

\begin{center}
A. Ballesteros$^1$, E. Celeghini$^2$  and M.A. del Olmo$^3$
\end{center}

\begin{center}
$^1${\sl Departamento de F\'{\i}sica, Universidad de Burgos, \\
E-09006, Burgos, Spain.}\\
\medskip

$^2${\sl Departimento di Fisica, Universit\`a  di Firenze and
INFN--Sezione di
Firenze \\
I50019 Sesto Fiorentino,  Firenze, Italy}\\
\medskip

$^3${\sl Departamento de F\'{\i}sica Te\'orica, Universidad de
Valladolid, \\
E-47005, Valladolid, Spain.}\\
\medskip

{e-mail: angelb@ubu.es, celeghini@fi.infn.it, olmo@fta.uva.es}
\end{center}

\bigskip
\centerline{\today}
\bigskip

\begin{abstract}
The two isomorphic Borel subalgebras of $gl(n)$, realized on
upper and lower triangular matrices, allow us to consider the $gl(n)\oplus t_n$
algebra as a self-dual Drinfeld double.
Compatibility conditions impose the choice of an
orthonormal basis in the Cartan subalgebra and fix the basis of $gl(n)$. A natural  Lie
bialgebra structure on
$gl(n)$ is obtained, that offers a new perspective for its standard quantum deformation.  
\end{abstract}
\vskip 1cm

\vskip 0.4cm

\vfill
\eject

\section{Introduction\label{introduccion}}

Drinfeld-Jimbo deformations of semisimple Lie algebras \cite{Dri85,Jimbo} are closely related to quantum
doubles
\cite{Dri87}, and their universal quantum  $R$-matrices can be computed by making use of this property 
(see, for instance, Ref.~\cite{Dri87}-\cite{CP}). For a complete discussion
of the problem we refer to~\cite{CP}, where it is shown that 
$U_h(g)$ (where $g$ is a finite dimensional complex semisimple Lie algebra)
 is `almost' a quantum double. However, the positive and negative quantum Borel subalgebras have
in common the Cartan subalgebra, and the underlying Drinfeld double (Manin triple) cannot be properly
defined.

In this paper we propose a way to circumvent this problem by enlarging the Cartan subalgebra in such a way
that two disjoint solvable algebras, isomorphic to  Borel subalgebras, can be properly paired.
Therefore, a Manin triple structure for the extended algebra  $\bar g=g\oplus
t$, where $t$ is an abelian algebra, is explicitly constructed and analyzed.

As a first example of this approach, we consider the $gl(n)$ case. We introduce the extended algebra 
$\bar g=gl(n)\oplus t_n$, where
$t_n$ is an
$n$-dimensional abelian algebra generated by $I_j, (j=1,\dots,n)$. 
Explicitly, we define the following new basis in the $2n$ dimensional abelian subalgebra of $\bar g$:
\be
H_j^+=\frac{1}{\sqrt{2}}(H_j + i I_j),
\qquad\qquad H_j^-=\frac{1}{\sqrt{2}}(H_j - i I_j)
\label{hi}
\ee
Afterwards, we consider two disjoint solvable Lie algebras $s_+$ and $s_-$, that contain the $H_j^+$ and
$H_j^-$ generators, respectively, and that are isomorphic to  the subalgebras defined by upper and lower
triangular  matrices of
$gl(n)$. Then, the definition of the pairing is done in terms of the Killing form for the $gl(n)$
generators and it is generalized through an appropriate definition
for the additional central generators.
As a result the Manin triple \cite{Dri87,majid,CP} $(s_+,s_-,\bar g)$  is
obtained.

This result offers a way for the quantization of  $gl(n)\oplus t_n$.
From the underlying Lie bialgebra we can deduce that the direct sum structure is no longer preserved after
deformation, since the extra abelian sector gives rise to a set of twists that will intertwine with the
standard quantization of the
$gl(n)$ subalgebra (results concerning the known $gl(n)$ quantizations can be found in
Ref.~\cite{Sudbery}-\cite{FGgln}).
Moreover, from $U_z(gl(n)\oplus t_n)$ 
one would be able to recover
$U_z(sl(n))$ in the representation of $t_n$ in which all $I_i$ are equal.
Note also that while this ``central extension procedure" (\ref{hi}) can be only done on the complex,
the final
$U_z(sl(n))$ would be obtained on $\R$.

The paper is organized as follows. In section 2 we recall the basic notation. Section 3 is devoted to the
$gl(2)$ case, that is fully discussed. The generalization to $gl(n)$ is presented in section 4 and some
comments and remarks are included in section 5.


\section{Manin triples and Drinfeld doubles}

In order to fix the notation, let us recall that a Manin triple  is a set of three
Lie algebras
$(s_+,s_-,\bar g)$ that can be constructed as follows. 

Firstly, let us consider a pair of disjoint Lie algebras
$s_+$ and
$s_-$ with the same dimension. Their commutation rules are given by the structure tensors $f$ and
$c$:
\be
[Z_p,Z_q]= f^r_{p,q}\,Z_r, \qquad Z_p, Z_q\in s_+ ,
\label{efes}
\ee
\be
[z^p,z^q]= c_r^{p,q}\,z^r ,\qquad z^p,z^q\in s_- .
\label{ces}
\ee
Now let us define a  pairing between them (i.e., a non-degenerate symmetric
bilinear form on the vector space $s_+ \oplus s_-$ for which $s_\pm$ are
isotropic)
\be
\langle Z_p,Z_q\rangle=0  ,\qquad
\langle Z_p,z^q\rangle=\delta_p^q , \qquad
\langle z^p,z^q\rangle=0  .
\label{pairingz}
\ee
If the compatibility relations (crossed Jacobi identities) 
 \be\label{compatibility}
c^{p,q}_r f^r_{s,t} = c^{p,r}_s f^q_{r,t}+c^{r,q}_s f^p_{r,t}
+c^{p,r}_t f^q_{s,r} +c^{r,q}_t f^p_{s,r}  \; 
\ee  
are fulfilled, we can construct a new Lie algebra $\bar g$ such that, as a vector space, $\bar g=s_+ \oplus
s_-$ and such that the pairing is invariant under the adjoint representation of $\bar g$ (i.e., 
$\langle [a,b],c\rangle=-\langle a,[b,c]\rangle],\; \forall a,b,c \in \bar g$). The latter
condition leads to the following crossed commutation rules between the elements of $s_+$ and $s_-$: 
\be
[z^p,Z_q]= f^p_{q,r}z^r- c^{p,r}_q Z_r .
\label{zz}
\ee

The Lie algebra $\bar g$ is then
called a Drinfeld double \cite{Dri87}, and $(s_+,s_-,\bar g)$ is called a Manin triple. Then
$\bar g$  can be endowed with a (quasitriangular) Lie bialgebra structure $(\bar g,\delta_{D})$ 
\be
\delta_{D}(Z_p)=-\eta(Z_p)=- c_p^{q,r}\,Z_q\otimes Z_r ,\qquad 
\delta_{D}(z^p)=\delta(z^p)=f^p_{q,r}\,z^q\otimes z^r .
\label{deltaD}
\ee
This ``double Lie bialgebra" has as Lie sub-bialgebras   
  ($s_+,-\eta$) and its dual ($s_-,\delta$).
Obviously,
several Manin triple structures for a given $\bar g$ can be constructed (see, for instance,
Ref.~\cite{BelDr}-\cite{6dim}). 

Finally, we remark that (\ref{deltaD}) can be derived either from the classical
$r$-matrix
\[ 
r=\sum_p{z^p\otimes Z_p},
\]
or from its skew-symmetric counterpart 
\be \tilde r=\frac12 \sum_p{z^p\wedge Z_p}  \;.
\label{rmatt}
\ee


\section{The Drinfeld double $gl(2)\oplus t_2$}

Let us start with the elementary example of $gl(2)$ by considering the solvable algebras 
$ 
s^+=\{Z_1,Z_2,Z_{3}\}
$
and 
$
s^-=\{z^1,z^2,z^{3}\}
$
with commutation
rules 
\be\label{conmutacionesesemas}
[Z_1,Z_2]=0 ,\qquad
[Z_1,Z_{3}]=\frac{1}{\sqrt{2}} Z_{3}, \qquad
[Z_2,Z_{3}]=-\frac{1}{\sqrt{2}} Z_{3}, 
\ee
\be\label{conmutacionesesemenos}
[z^1,z^2]=0 ,\qquad
[z^1,z^{3}]=-\frac{1}{\sqrt{2}} z^{3}, \qquad
[z^2,z^{3}]=\frac{1}{\sqrt{2}} z^{3}. 
\label{smenos}
\ee
The structure tensors for $s_+$, $f^p_{q,r}$ (\ref{efes}), and $s_-$, $c^{p,r}_q$ (\ref{ces}), 
are
\[
f^{3}_{1,3}=-f^{3}_{3,1}=\frac{1}{\sqrt{2}} ,\qquad
f^{3}_{2,3}=-f^{3}_{3,2}=-\frac{1}{\sqrt{2}},\qquad 
c^{p,q}_{r}=-f^{r}_{p,q} .
\label{tensores}
\]

Now, let us consider the triple $(s_+,s_-,\bar g=s_+ + s_-)$ and a
bilinear form on $\bar g$ defined through (\ref{pairingz}). Jacobi identities
(\ref{compatibility}) are easily checked  and the  crossed commutation rules between $s_+$ and $s_-$ are
given by (\ref{zz}):
\be\begin{array}{l}\label{conmutacionescrossed}
 [z^1,Z_{3}]=-[z^2,Z_{3}]=\frac{1}{\sqrt{2}} \,Z_{3},\\[0.3cm]
 [z^{3},Z_{1}]=-[z^{3},Z_{2}]=\frac{1}{\sqrt{2}} \,z^{3},\\[0.3cm]
[z^{3},Z_{3}]=-\frac{1}{\sqrt{2}} (z^1+Z_1) +
\frac{1}{\sqrt{2}} (z^2+Z_2).
\end{array}\ee
Note that the $1/\sqrt{2}$ factor in
(\ref{conmutacionesesemas}) and (\ref{conmutacionesesemenos}) is essential in this construction. Since
$s_+$ and
$s_-$ are isomorphic, we are dealing with a self-dual Manin triple. In other words, ($s_+,\eta$) is a
Lie bialgebra with cocommutator
$$ \eta(Z_p)= c^{q,r}_p Z_q\otimes Z_r .
\;  
\label{agc}
$$ 
Explicitly:
$$
\eta(Z_1)=\eta(Z_2)=0 , \qquad \eta(Z_3)=\frac{1}{\sqrt{2}}Z_3\wedge  (Z_1 - Z_2) .
$$
Respectively, ($s_-,\delta$) is the dual
Lie bialgebra with cocommutator
$$ 
\delta(z^p)=f^{p}_{q,r} z^q\otimes z^r ,
\label{aga}
$$ 
which reads
$$
\delta(z^1)=\delta(z^2)=0 , \qquad 
\delta(z^3)=-\frac{1}{\sqrt{2}}z^3 \wedge (z^1 - z^2) .
$$

Now,  considering the change of basis
\be\begin{array}{lll}\label{baseDD}
& H_1=\frac{1}{\sqrt{2}}(Z_1 + z^1) ,\qquad
&I_1=\frac{1}{i\sqrt{2}}(Z_1 - z^1), \\[0.3cm]
& H_2=\frac{1}{\sqrt{2}}(Z_2 + z^2),\qquad
& I_2=\frac{1}{i\sqrt{2}}(Z_2 - z^2),\\[0.3cm]
& F_{12}=Z_{3},\qquad
& F_{21}=z^{3},
\end{array}\ee
and  rewriting the relations
(\ref{conmutacionesesemas}), (\ref{conmutacionesesemenos})
and(\ref{conmutacionescrossed}), 
we obtain
\be\begin{array}{lllll}\label{conmutaciones gldos}
& [I_i, \cdot]=0,\quad & [H_1,H_2]=0, \quad
&[H_1,F_{12}]=F_{12}, \quad
&[H_1,F_{21}]=-F_{21}, \\[0.3cm]
&[H_2,F_{12}]=-F_{12}, \quad
&[H_2,F_{21}]=F_{21}, \quad
&[F_{12},F_{21}]=H_1-H_2,&
\end{array}\ee
which are just the commutation rules for the Lie algebra
$\bar g=gl(2)\oplus t_2$ in the usual basis
$\{H_1,H_2, F_{12}, F_{21} \}
\oplus
\{I_1,I_2 \}$. 

Therefore, we have proven that the two solvable algebras  $s_+$ and $s_-$
together with the pairing (\ref{pairingz}) endow $\bar g=gl(2)\oplus t_2$ with a
Drinfeld double structure.   Note that $s_+$ and
$s_-$ have been chosen to be isomorphic to the upper and lower triangular $2\times 2$ matrices
 of 
$gl(2)$, respectively.

Explicitly, the associated
Lie bialgebra (\ref{deltaD}) is
\[\begin{array}{l}\label{bialgebra gldos}
\delta_{D} (I_i)=0, \\[0.3cm]  \delta_{D} (H_i)=0,\\[0.3cm]
\delta_{D} (F_{12})=-\frac 12 F_ {12} \wedge (H_1-H_2)- \frac i2 F_ {12}
\wedge (I_1-I_2), \\[0.3cm]
\delta_{D} (F_{21})=-\frac 12 F_ {21} \wedge (H_1-H_2)+\frac i2 F_ {21}
\wedge (I_1-I_2).
\end{array}\]
This cocommutator $\delta_D$ can be derived from the classical $r$-matrix
(\ref{rmatt}) i.e.
\[ \tilde r=
\frac12 \, F_{21}\wedge F_{12} 
+ \frac{i}{4}\left( H_{1}\wedge I_{1} + H_{2}\wedge I_{2}  \right) = \tilde r_s + \tilde r_t ,
\label{rmatt2}
\]
where $\tilde r_s$ generates the standard deformation of $gl(2)$ and $\tilde r_t$ denotes a twist,
that becomes trivial in the representation of $t_2$ where $I_1-I_2=0$.

It is worth to note that from  (\ref{baseDD})  the pairing (\ref{pairingz})  is simply   the
Killing form relations for $gl(2)$ in the ``oscillator representation" convention
\cite{Sciarrino}
\[\label{pairing21}
\langle H_{i},F_{jk}\rangle=0,\qquad  \langle H_i,H_j\rangle=\delta_{ij}, \qquad
\langle F_{ij},F_{kl}\rangle=\delta_{jk}\,\delta_{il} 
\]
supplemented by a suitable definition of the pairing for the
additional central generators
\[\label{pairing22}
\langle I_{i},I_{j}\rangle=\delta_{ij},\qquad  \langle I_i,H_j\rangle=0, \qquad
\langle I_i,F_{jk}\rangle=0.
\]

\section{The $gl(n)\oplus t_n$ case}

For the general case of $gl(n)$ the procedure is similar to the one followed in the preceding case
of  
$gl(2)$. We consider two $n(n+1)/2$-dimensional solvable Lie algebras $s_+$ and $s_-$, isomorphic to 
the subalgebras defined by upper and lower triangular  $n\times n$ matrices of $gl(n)$,
 with generators
\[\begin{array}{llll}
s^+:  &\{ X_i,Y_{ij}\},&\qquad & i,j=1,\dots,n,\quad i<j,
\\[0.3cm]
s^-:  &\{ x^i,y^{ij}\},&\qquad & i,j=1,\dots,n,\quad i<j ,
\end{array}\]
and commutation rules given by
\[\begin{array}{llll}\label{snmas}
& [X_i,X_j]=0, \qquad
& [X_i,Y_{jk}]=\frac{1}{\sqrt{2}}(\delta_{ij} - \delta_{ik})\, Y_{jk},\qquad
& [Y_{ij},Y_{kl}]=(\delta_{jk} Y_{il}- \delta_{il} Y_{kj}) ,\\[0.3cm]
& [x^i,x^j]=0, \qquad
& [x^i,y^{jk}]=- \frac{1}{\sqrt{2}}(\delta_{ij} - \delta_{ik})\, y^{jk},\qquad
& [y^{ij},y^{kl}]=- (\delta_{jk} y^{il}- \delta_{il} y^{kj}).
\end{array}\]
Following (\ref{efes}) and (\ref{ces}), the corresponding structure tensors $f$ and $c$ read
\[\begin{array}{lll}
f_{i,jk}^{lm}=-f_{jk,i}^{lm}=-c^{i,jk}_{lm}=c^{jk,i}_{lm}=
\frac{1}{\sqrt{2}}(\delta_{ij}-\delta_{ik})\delta_{jl}\delta_{km},\\[0.3cm]
f_{ij,kl}^{mn}=-c^{ij,kl}_{mn}=\delta_{jk}\delta_{im}\delta_{ln} -
\delta_{li}\delta_{km}\delta_{jn} .
\end{array}\]

If we assume that the two algebras are paired by
\be\label{pairing}
\langle x^i,X_j\rangle= \delta^i_j ,\qquad 
\langle y^{ij},Y_{kl}\rangle= \delta^i_k\delta^j_l ,
\ee
we can define a bilinear form on the vector space $s_+ \oplus s_-$ in terms of
(\ref{pairing}) such that both $s_\pm$ are isotropic for it.  Under these
conditions we can consider the triple
$(s_+,s_-,\bar g=s_+ + s_-)$. Indeed, by taking into account (\ref{zz}) we obtain the crossed commutation
rules 
\[\begin{array}{l}\label{crossed1}
 [x^i, X_j] = 0 ,\qquad
[x^i, Y_{jk}]\; =\;\frac{1}{\sqrt{2}} (\delta_{ij} - \delta_{ik})\; Y_{jk},\qquad
[y^{ij}, X_k]\; =\; \frac{1}{\sqrt{2}} (\delta_{ik} - \delta_{jk})\; y^{ij},\\[0.3cm]
 [y^{ij}, Y_{kl}] = \,\left\{\delta_{ik} (Y_{jl} + y^{lj})-\delta_{jl} 
(Y_{ki} + y^{ik})\right\} - \delta_{ki}\, \delta_{lj}\,(X_i + x^i - X_j- x^j ),
\label{crossed2}
\end{array}\]
where $Y_{ij}\equiv 0$ and   $y^{ij} \equiv 0$ for $i>j$.

We can avoid to check the compatibility conditions (\ref{compatibility}) since $\bar g$ is a well known
Lie algebra, that can be identified by considering the following change of basis
\[
H_i=\frac{1}{\sqrt{2}}(X_i + x^i), \qquad
I_i=\frac{1}{i\sqrt{2}}(X_i - x^i) ,\qquad
F_{ij}=Y_{ij}+y^{ji}.
\label{change}
\]
The full set of commutation rules for $\bar g$ is:
\[\begin{array}{l}
[I_i,\cdot]=0 ,\qquad [H_i,H_j]=0, \qquad
[H_i,F_{jk}]=(\delta_{ij} - \delta_{ik})F_{jk}, \\[0.3cm]
[F_{ij},F_{kl}]=(\delta_{jk} F_{il}- \delta_{il} F_{kj}) + \delta_{jk} 
\delta_{il}(H_i - H_j) . 
\end{array}\]
These relations are nothing but the commutation rules for $gl(n)\oplus t_n$, where
$gl(n)$ is defined by the usual fundamental representation in terms of the
$n\times n$ matrices
$H_i$  and $F_{ij}\; (i\neq j)$ defined as follows 
\[\label{matrices}
(H_i)^{jk}=\delta_i^j\,\delta_i^k ,\qquad
(F_{ij})^{kl}=\delta_i^k\,\delta_j^l  ,\qquad i,j,k,l=1,\dots,n.
\]
 Thus,
$(s_+,s_-,\bar g=gl(n)\oplus t_n)$ is a self-dual Manin triple.

The cocommutator $\delta_{D}$  (\ref{deltaD}) reads
\[\begin{array}{ll}\label{bialgebra gln}
\delta_{D} (I_i)=0, \\[0.3cm]
  \delta_{D} (H_i)=0,\\[0.3cm]
\delta_{D} (F_{ij})=-\frac 12 F_{ij} \wedge (H_i-H_j)- \frac i2 F_{ij}
\wedge (I_i-I_j) + \sum_{k=i+1}^{j-1}{F_{ik} \wedge F_{kj}}, \qquad & i<j ,\\[0.3cm]
\delta_{D} (F_{ij})=\frac 12 F_{ij} \wedge (H_i-H_j)- \frac i2 F_{ij}
\wedge (I_i-I_j) - \sum_{k=j+1}^{i-1}{F_{ik} \wedge F_{kj}},\qquad & i>j.
\end{array}\]
 In particular, ($s_+,\delta_{D}$) and its dual ($s_-,\delta_{D}$) are Lie sub-bialgebras. 
The classical $r$-matrix  (\ref{rmatt}),  in the basis
$\{H_i,F_{ij} ,I_i \}$ , 
is written as
\[ \tilde r= 
\frac12 \, \sum_{i<j}{F_{ji}\wedge F_{ij} }
+ \frac{i}{4}\sum_{i}{H_{i}\wedge I_{i} } = \tilde r_s + \tilde r_t .
\label{rmatt3}
\]
Again $\tilde r_s$ generates the standard deformation of $gl(n)$ and 
$\tilde r_t$ is a twist
(not of Reshetikhin type \cite{reshetikhin}).
When all the $I_i$ are equal the twist
$\tilde r_t$ becomes trivial. 

Note also that the chain
$gl(m)\subset gl(m+1)$ is preserved at the level of Lie bialgebras.


\section{Concluding remarks}

We have introduced a Drinfeld double structure on the Lie algebra $\bar g=gl(n)\oplus t_n$, 
inspired by the Cartan-Dynkin approach for the classification of simple Lie algebras in which the
(solvable) Borel subalgebras play an essential role. We briefly comment on some consequences of this
construction both from a classical viewpoint and from the quantum deformation perspective.

Firstly, it is worth
to emphasize that the choice of the basis in $\bar g$ is crucial because in order to have a Drinfeld
double structure we need, up to a global factor, both an orthonormal basis in the Cartan subalgebra and a
fixed normalization for all the remaining generators.
In terms of the associated Killing form $K$, this choice implies
that for the ``extended" Cartan subalgebra
$K_{mn}=\delta_{m,n}$ while 
$K_{\alpha\beta}=\delta_{\alpha,-\beta}$
for the root vectors. This seems to be the natural choice and corresponds to the ``oscillator
representation" convention
\cite{Sciarrino} (many others have been considered in the
literature \cite{Cornwell}).

Secondly, the results here presented can be thought of as a first step
in order to approach quantum deformations of semisimple Lie algebras from a quantum double
perspective.
In particular, we have 
provided the first order information for the construction of  $U_z(gl(n)\oplus t_n)$  and, as a
byproduct, of the Drinfeld-Jimbo deformation $U_z(sl(n))$. Under this approach the essential task is to
obtain the full quantization of a solvable Lie algebra in which the only deformed coproducts correspond
to the nilpotent generators (one could follow Ref.~\cite{6dim} and \cite{3dim}). In this way, Serre
relations do not play any role since all the root vectors are considered on the same footing, and the
quantization procedure is rather simplified by the self-dual nature of the underlying Lie bialgebra (and
of the associated Poisson-Lie group). 

By following the same lines, the discussion of the Drinfeld double structure for the $B,C$ and $D$ series
of simple Lie algebras will be presented elsewhere.

\small

\section*{Acknowledgments}

This work was partially supported  by the Ministerio de
Educaci\'on y Ciencia  of Spain (Projects BMF2002-02000, FIS2005-02000 and FIS2004-07913),  by the
Junta de Castilla y Le\'on   (Project VA013C05), and by
INFN-CICyT (Italy-Spain).



\begin{thebibliography}{99}

\bibitem{Dri85} V.G. Drinfeld, {\it Dokl. Akad. Nauk. SSSR} {\bf 283} (1985), 1060

\bibitem{Jimbo} M. Jimbo,
{\it Lett. Math. Phys}.  {\bf 10} (1985) 63   

\bibitem{Dri87} V.G. Drinfeld, ``Quantum Groups'' in {\it Proceedings of
the International Congress of Mathematicians}, Berkeley, 1986, A.M. Gleason (ed.)   
(AMS, Providence, 1987)

\bibitem{Burr} N. Burroughs,  {\it Comm. Math. Phys.} {\bf 127} (1990) 109

\bibitem{FG} C. Fronsdal and A. Galindo,
{\it Lett. Math. Phys}.  {\bf 27} (1993) 59   

\bibitem{Vladimirov} A.A. Vladimirov,
{\it Mod. Phys. Lett A}.  {\bf 8} (1993) 2573   

\bibitem{majid}  S. Majid, {\em Foundations of Quantum
Group Theory}  (Cambridge Univ. Press, Cambridge,  1995)

\bibitem{CP}  V. Chari and A. Pressley, {\em A Guide to Quantum Groups},
(Cambridge Univ. Press, Cambridge 1994)

\bibitem{Sudbery} A. Sudbery,
  {\it J. Phys. A: Math. Gen.} {\bf 23} (1990) L697

\bibitem{Schir} A. Schirrmacher,
  {\it Z. Phys. C} {\bf 50} (1991) 321

\bibitem{reshetikhin} N. Yu Reshetikhin, {\it Lett. Math. Phys.} {\bf 20} (1990) 331

\bibitem{Alisauskas} S. Alisauskas and Yu F. Smirnov,
  {\it J. Phys. A: Math. Gen.} {\bf 27} (1994) 5925

\bibitem{Dobrev} V.K. Dobrev and P. Parashar,
  {\it J. Phys. A: Math. Gen.} {\bf 26} (1993) 6991;\\
   {\it J. Phys. A: Math. Gen.} {\bf 32} (1999) 443

\bibitem{FGgln} C. Fronsdal and A. Galindo,
{\it Lett. Math. Phys}.  {\bf 34} (1995) 25   

\bibitem{BelDr} A.A. Belavin and V.G. Drinfeld, {\it Funct. Anal. Appl.} {\bf 16} (1983) 159

\bibitem{gomez} X. Gomez, {\it J. Math. Phys}. {\bf 41} 
(2000) 4939

\bibitem{HS}
L. Hlavaty and L. Snobl, {\it Int. J. Mod. Phys.}  {\bf A17}
(2002) 4043    \\
L. Snobl, {\it J. High Energy Phys.}  {\bf 9}
(2002) Art. 018   

\bibitem{6dim} A. Ballesteros, E. Celeghini and  M.A. del Olmo,
  {\it J. Phys. A: Math. Gen.} {\bf 38} (2005) 3909

\bibitem{Sciarrino}  L. Frappat, A. Sciarrino and P. Sorba, {\em Dictionary on Lie Algebras and
Superalgebras}  (Academic Press, London,  2000)

\bibitem{Cornwell}  J.F. Cornwell, {\em Group Theory in Physics},
(Academic Press, London 1984)

\bibitem{3dim} A. Ballesteros, E. Celeghini and  M.A. del Olmo,
  {\it J. Phys. A: Math. Gen.} {\bf 37} (2004) 1



\end{thebibliography}
\end{document}